\documentclass{mn2e}
\usepackage{epsf}

\newif\ifAMStwofonts

\def\gtorder{\mathrel{\raise.3ex\hbox{$>$}\mkern-14mu
             \lower0.6ex\hbox{$\sim$}}}
\def\ltorder{\mathrel{\raise.3ex\hbox{$<$}\mkern-14mu
             \lower0.6ex\hbox{$\sim$}}}

\title[SDSS J120923.7$+$264047]{SDSS J120923.7$+$264047: A new massive galaxy cluster with a bright giant arc}

\author[E.~O.~Ofek et al.]
       {Eran~O.~Ofek\thanks{e-mail: eran@astro.caltech.edu}$^{1}$, Stella~Seitz$^{2}$, Felix Klein$^{2}$ \\
        $^{1}$ Division of Physics, Mathematics and Astronomy, California Institute of Technology, Pasadena, CA 91125 \\
        $^{2}$ Universit\"{a}ts-Sternwarte M\"{u}nchen, M\"{u}nchen, Germany}

\date{Accepted ?
      Received ?
      in original form ?}

\begin{document}

\maketitle

\begin{abstract}

Highly magnified lensed galaxies
allow us to probe 
the morphological and spectroscopic properties
of high-redshift stellar systems in great detail.
However, such objects are rare, and there are
only a handful of lensed galaxies which are bright
enough for a high-resolution spectroscopic study
with current instrumentation.
We report the discovery of a new massive lensing cluster,
SDSS\,J120923.7$+$264047, at $z=0.558$.
Present around the cluster core, at angular distances of up to $\sim40''$,
are many arcs and arc candidates,
presumably due to lensing of background galaxies by the
cluster gravitational potential.
One of the arcs, $21''$ long, has an $r$-band magnitude of 20, making
it one of the brightest known lensed galaxies.
We obtained a low-resolution spectrum of this galaxy, using the Keck-I telescope,
and found it is at redshift of $z=1.018$.

\end{abstract}


\section{Introduction}
\label{Introduction}

High redshift galaxies ($z\gtorder1$) are
out of reach for current high-resolution
spectrographs.
Fortunately, such galaxies may be highly magnified,
when lensed
by galaxies or cluster of galaxies, making them
bright enough for high-resolution spectroscopic studies.
%
%
However, high magnification lensed galaxies are rare,
and only small number of targets suitable 
for high-resolution spectroscopic studies are known.
Among these are:
MS1512$-$cB58 (Yee et al. 1996), with $r=20.4$\,mag at $z=2.72$;
an $r=20.3$\,mag Lyman break galaxy at $z=3.07$ (Smail et al. 2007);
and finally a highly magnified galaxy, at $z=2.73$,
with $r=19.2$\,mag (Allam et al. 2007).
%
%
Studies of these objects
(e.g., Teplitz et al. 2000; Pettini et al. 2000; 2002; Baker et al. 2004; Schaerer \& Verhamme 2008)
provided valuable information
on their metallicity and inter-stellar medium.

In this paper we report on the discovery of a previously unknown
rich cluster of galaxies at $z=0.558$.
The cluster containing
a bright ``giant arc'', which is an appropriate target
for high-resolution spectroscopic studies.
In addition, the cluster, which has a considerable Einstein radius
of probably $\sim20-40''$, contains a large number of additional
arcs and lensed galaxies. This cluster is probably
as rich as the Abell 1689 lensing cluster (Broadhurst et al. 2005),
and therefore it is an excellent target for detailed
strong lensing and weak lensing modeling (e.g., Brada\u{c} et al. 2006; Halkola et al. 2006).

The outline of this paper is as follows.
We describe the observations in \S\ref{Obs} and 
in \S\ref{Arcs} we discuss the cluster and arcs
along with their measured properties.
Finally, we briefly discuss the results in \S\ref{Sum}.

\section{Observations}
\label{Obs}

The discovery of the cluster and arcs described in this paper,
was made as part of a survey for large separation
lensed quasars (e.g., Maoz et al. 1997; Ofek et al. 2001; Phillips et al. 2001;
Inada et al. 2003; Oguri et al. 2004; Inada et al. 2006; Oguri et al. 2008),
among Sloan Digital Sky Survey (SDSS; York et al. 2000) photometrically selected quasars.
This search will be described elsewhere.
Our search uncovered a previously unknown cluster of galaxies
in an SDSS image.
In addition to the cluster,
a giant arc was clearly visible near the cluster,
in the relatively shallow SDSS images.

\begin{figure*}
\centerline{\epsfxsize=15cm\epsfbox{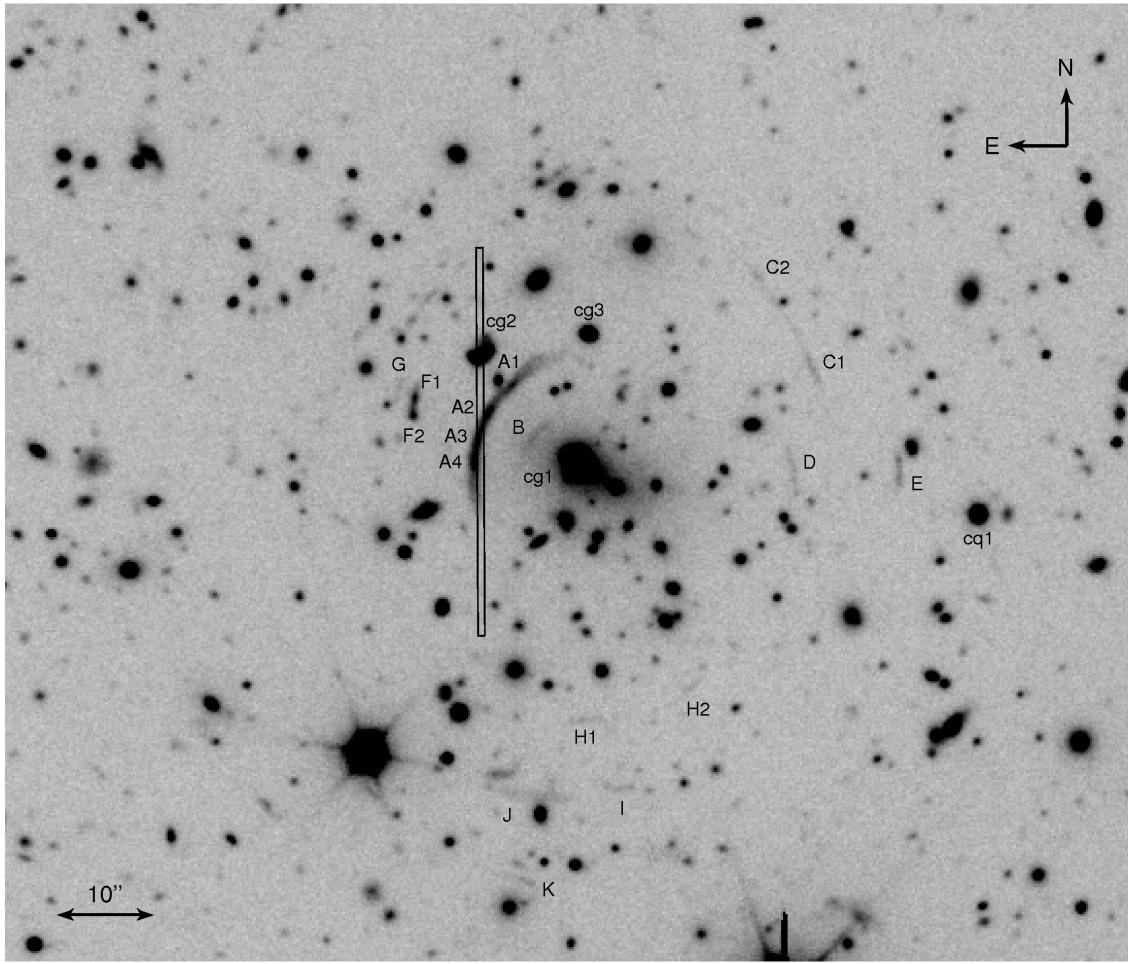}}
\caption{$R$-band image of SDSS\,J120923.7$+$264047,
obtained with LRIS mounted on the Keck-I 10\,m telescope.
Objects labeled by A1..K are candidate arcs.
The properties of the labeled arc candidates are listed
in Table~\ref{Tab-Pos}.
Also labeled are: the cluster bright galaxy
(cg1; SDSS\,J120923.68$+$264046.7) which is at $z=0.558$;
(cg2; SDSS\,J120924.41$+$264057.8) at $z=0.542$;
(cg3; SDSS\,J120923.60$+$264059.9) at $z=0.542$;
and a bright SDSS quasar (cq1; SDSS\,J120920.61$+$264041.2),
at $z=1.555$.
The quasar, cq1, which has an $r$-band magnitude of 19.2 is found
$41''$ from the bright cluster galaxy, so most probably
it is not strongly lensed by the cluster.
Also marked is the position (and actual width) of the slit
we used to obtain the spectrum of the arc.}
\label{SDSS1209R}
\end{figure*}

\begin{figure}
\centerline{\epsfxsize=8cm\epsfbox{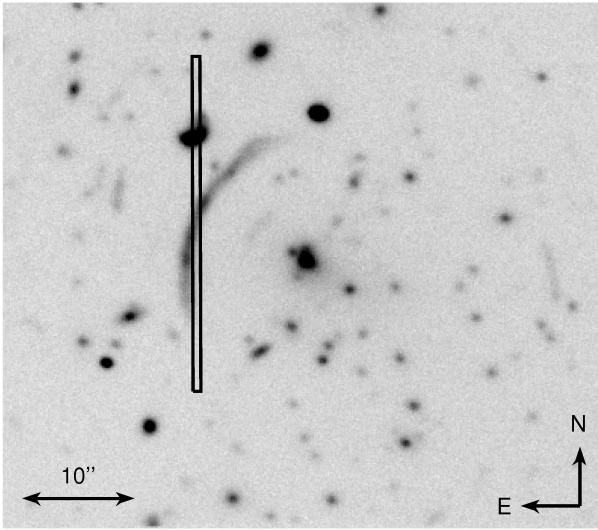}}
\caption{$g$-band image of SDSS\,J120923.7$+$264047,
obtained with LRIS mounted on the Keck-I 10\,m telescope.
See Fig.~\ref{SDSS1209R} for details.}
\label{SDSS1209g}
\end{figure}

On UTC 2008 Jan 04, we obtained $5\times220$\,s $g$-band
and $5\times180$\,s $R$-band images of the cluster environment using
the Low Resolution Imaging Spectrograph (LRIS; Oke et al. 1995)
on the Keck-I telescope, under seeing of $0.6''$.
The combined Keck $R$-band image is presented in Figure~\ref{SDSS1209R},
and the combined $g$-band image is presented in Fig.~\ref{SDSS1209g}.
The images were calibrated astrometrically using six objects
appearing in both the Keck and SDSS images.
The solution root-mean-square (RMS) is
better than $0.1''$ in both axes and both images.
In Table~\ref{Tab-Pos} we give the positions and photometry
of the main arcs detected in the images.
We have measured the magnitude of each arc, by
combining all the light within a polygon
defined by the SExtractor segmentation (Bertin \& Arnouts 1996).
In case SExtractor broke an arc to several segments,
we combined the light within these segments.
The SExtractor detection threshold was set to
about 2-$\sigma$ above background.

\begin{figure}
\centerline{\epsfxsize=8cm\epsfbox{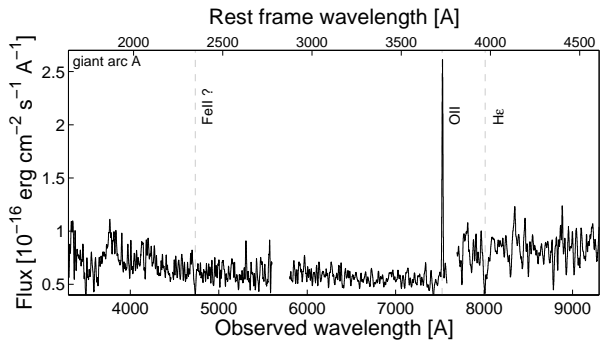}}
\caption{The Keck/LRIS spectrum of the giant arc A
(Fig.~\ref{SDSS1209R}; Tab.~\ref{Tab-Pos}).
The detection of several lines indicates a redshift of 1.018.}
\label{GiantArcSpec}
\end{figure}

\begin{figure}
\centerline{\epsfxsize=8cm\epsfbox{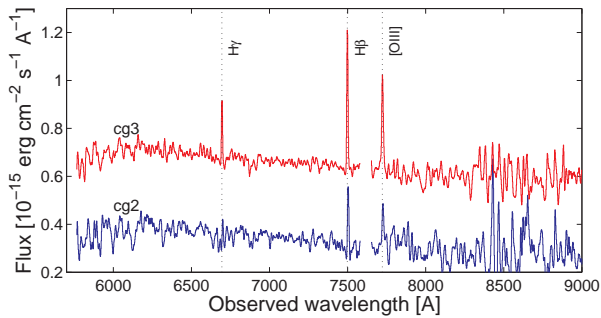}}
\caption{The Keck/LRIS spectrum of the galaxies cg2 and cg3.
The galaxies are at redshift of $0.542$.
For clarity, the spectrum of cg3 is shifted upward by $4\times10^{-16}$\,erg\,cm$^{-2}$\,s$^{-1}$\,\AA$^{-1}$.}
\label{Spec_2blue_obj}
\end{figure}

On the same night, using LRIS--Atmospheric Dispersion Compensator we also obtained a 1300\,s spectrum of
the bright
giant arc, labeled A1...A4 in Fig.~\ref{SDSS1209R}.
We used a $0.7''$ slit with the 5600\,\AA~dichroic;
the 400/3400 grism on the blue side; and
the 400/8500 grating, centered on $7693$\,\AA, on the red arm.
The slit position angle was set to $180.2$\,deg, in
order to include also the blue (SDSS $g-r\approx0.6$)
galaxy labeled cg2 in Fig.~\ref{SDSS1209R}.
The spectrum was flux calibrated using observations,
obtained on the same night,
of the spectrophotometric standard Hz2 (Turnshek et al. 1990).
The spectrum of arc A is shown in Fig.~\ref{GiantArcSpec}.
We also obtained using the same setup, on UTC 2007 Dec 15,
a 1500\,s-exposure spectra of the galaxies cg2 and cg3.
The spectral reduction was done using tools in
the MATLAB environment (Ofek et al. 2006).
The spectra of the galaxies cg2 and cg3 are shown in Fig.~\ref{Spec_2blue_obj}.

\section{The cluster and arcs}
\label{Arcs}

The Keck $R$-band image of the cluster is presented in Fig.~\ref{SDSS1209R}.
In this figure, it is possible to identify
many arc candidates, most of which
are labeled (A to K).
We list the properties of these arcs and arc candidates
in Table~\ref{Tab-Pos}.
The most prominent arc in this system, labeled
as A1--A4 in Fig.~\ref{SDSS1209R}, is very large
and bright. This arc is located about $11''$ East
of the brightest cluster galaxy, and its length
is $21''$.
The spectrum of the arc, presented in Fig.~\ref{GiantArcSpec},
shows a single strong emission line and 
several absorption features, at a probable redshift of $z=1.018$.
We note that there is also an indication for a Balmer decrement
around 7700\,\AA~(observed wavelength).
The redshift is based on the detection
of a single emission line.
Therefore, we attempted to look for other possible explanations
for this emission line, and
we did~not find any other reasonable redshift solution.
Moreover, the Einstein radius increases with source redshift.
Therefore, the relatively small Einstein radius of arc-A
implies that it is at relatively low redshift.
Its redshift,
which places it in the background of the cluster,
along with its unique appearance,
confirms the nature of this arc as a lensed galaxy.

The~Einstein~radius of a system scales like
$[D_{ls}/(D_{l}D_{s})]^{1/2}$, where
$D_{ls}$, $D_{l}$ and $D_{s}$ are
the angular diameter distances between
the lens and source, observer and lens,
and observer and source, respectively
(e.g., Narayan \& Bartelmann 1996).
Therefore, for the system discussed here,
we expect that the Einstein radius for
a $z\approx4$ system to be about $30\%$ larger
than for a $z=1$ system\footnote{Assuming 3rd year WMAP
cosmological parameters: $\Omega_{\Lambda}=0.716$, $\Omega_{m}=0.268$, H$_{0}=70.4$\,km\,s$^{-1}$\,Mpc$^{-1}$; Spergel et al. (2007).}.
This suggest that at least some of the arclets further away from the center of the cluster
are most probably strongly lensed background galaxies at higher redshifts.

The brightest cluster galaxy,
cg1 (SDSS\,J120923.69$+$264046.7)
in Fig.~\ref{SDSS1209R},
was observed spectroscopically by the SDSS, and we find it to be at a redshift of
0.558.
As mentioned before, we also obtained a spectrum
of the blue galaxies marked cg2
(SDSS\,J120924.41$+$264057.8)
and cg3 (SDSS\,J120923.60$+$264059.9)
in Fig.~\ref{SDSS1209R}.
We measured a redshift of 0.542 to these galaxies using
the H$\beta$ (4862.7\,\AA) and [OIII] (5008.2\,\AA)
emission lines present in their spectra.
Given the redshift of the brightest cluster galaxy,
we suggest that the cluster redshift is 0.558.
However, confirmation of this redshift
requires a redshift measurement for additional
galaxies in this region.

\begin{table}
\begin{center}
\begin{tabular}{lcccc}
\hline
Name & R.A. J2000  & Dec. J2000  &  g  &   r \\
     &             &             & mag & mag \\
\hline
A1   & 12:09:24.19 & +26:40:54.7 &     &     \\
A2   & 12:09:24.38 & +26:40:51.9 &     &     \\
A3   & 12:09:24.46 & +26:40:49.1 &     &     \\
A4   & 12:09:24.48 & +26:40:47.0 &     &     \\
A    &             &             &20.3 &20.0 \\
B    & 12:09:24.00 & +26:40:49.9 &     &24.0 \\
C1   & 12:09:21.91 & +26:40:56.9 &     &24.8 \\
C2   & 12:09:22.31 & +26:41:05.9 &     &24.9 \\
D    & 12:09:22.04 & +26:40:46.5 &     &24.5 \\
E    & 12:09:21.22 & +26:40:45.8 &     &24.2 \\
F1   & 12:09:24.94 & +26:40:53.5 &     &23.4 \\
F2   & 12:09:24.96 & +26:40:51.6 &     &23.7 \\
G    & 12:09:25.06 & +26:40:54.5 &     &25.2 \\
H1   & 12:09:23.65 & +26:40:20.2 &     &25.6 \\
H2   & 12:09:22.82 & +26:40:23.5 &     &26.0 \\
I    & 12:09:23.38 & +26:40:13.2 &     &25.3 \\
J    & 12:09:24.17 & +26:40:12.6 &     &24.5 \\
K    & 12:09:24.10 & +26:40:03.6 &     &24.7 \\
\hline
\end{tabular}
\caption{Positions and approximate magnitudes of most of
the arc candidates detected in the Keck $R$-band image (Fig.~\ref{SDSS1209R}).
The magnitudes are approximate, and have typical uncertainties of 0.3\,mag.
The astrometry was performed relative to the SDSS.}
\label{Tab-Pos}
\end{center}
\end{table}

\section{Discussion}
\label{Sum}

Strong and weak lensing are being used
to trace the mass distribution of galaxy clusters.
However, detailed modeling of the mass distribution
of the cores of galaxy clusters requires the identification
of multiple images of lensed sources.
Such an identification needs good multi-band color information,
or/and good spatial resolution (e.g., Broadhusrt et al. 2005;
Sharon et al. 2006). Indeed we are planing additional
multi-band observations of this cluster,
which will enable a detailed modeling of
the cluster gravitational potential.

\begin{figure}
\centerline{\epsfxsize=8cm\epsfbox{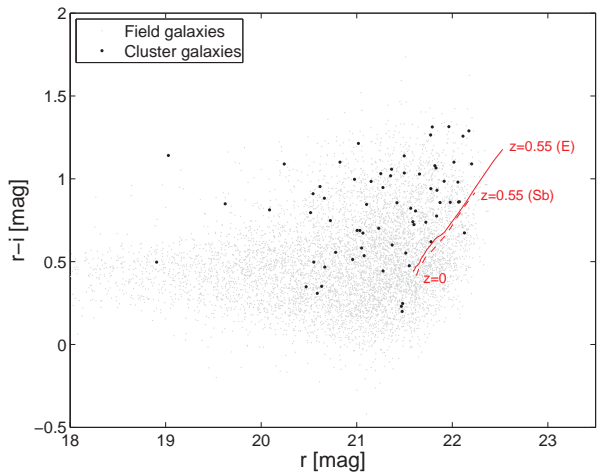}}
\caption{The color-magnitude diagram of SDSS galaxies
in the direction of the cluster.
The black circles represent
galaxies found with $2'$ from the cluster bright galaxy.
For comparison, we also show all the SDSS galaxies found
between $5'$ and $1^{\circ}$ from the cluster bright galaxy (gray dots).
The solid (dashed) red-lines display the locus of Elliptical (Sb)
galaxies as a function of redshift.
These lines take into account the Galactic extinction
($E_{B-V}=0.019$\,mag) in the direction of the
cluster (Schlegel et al. 1998; Cardelli et al. 1989).
The galaxies colors as a function of redshift
are based on synthetic photometry of galaxy
templates from Kinney et al. (1996).
The synthetic photometry was performed using
the code of Poznanski et al. (2002).}
\label{fig:SDSS1209_RedSequence}
\end{figure}

To confirm the presence of the cluster,
in Fig.~\ref{fig:SDSS1209_RedSequence} we show
the color magnitude diagram for SDSS galaxies
found with $2'$ from the cluster bright galaxy (black circles).
For comparison, we also show all the SDSS galaxies found
between $5'$ and $1^{\circ}$ from the cluster bright galaxy (gray dots).
The solid (dashed) red-lines display the locus of Elliptical (Sb)
galaxies as a function of redshift.
These lines take into account the Galactic extinction
($E_{B-V}=0.019$\,mag) in the direction of the
cluster (Schlegel et al. 1998; Cardelli et al. 1989).
It is evident, that in the cluster direction
there is an excess of high-redshift galaxies, relative to the field.

At the approximate position of the bright cluster galaxy,
we find a known ROSAT source (Voges et al. 2000).
Using the PIMMS
webtool\footnote{http://cxc.harvard.edu/toolkit/pimms.jsp},
we converted the ROSAT/PSPC count rate to an unabsorbed flux
of $6.9\times10^{-13}$\,erg\,cm$^{-2}$\,s$^{-1}$,
assuming a Galactic neutral Hydrogen column density of
$1.8\times10^{20}$\,cm$^{-2}$ (Dickey \& Lockman 1990; Kalberla et al. 2005)
and a power-law spectrum
with a photon index of $\Gamma=1$
(assuming thermal optically thin Bremsstrahlung).
Given the luminosity distance to the cluster, this flux implies
a luminosity of $\sim8\times10^{44}$\,erg\,s$^{-1}$ in the 0.1--2.4\,keV band.
Using the X-ray luminosity-mass relation given by
Reiprich \& B\"{o}hringer (2002), we estimate that the
$M_{200}$ mass\footnote{$M_{200}$ is the mass enclosed within $r_{200}$, which is the radius in which the mean density equals 200 times the mean density of the Universe (e.g., Navarro, Frenk, \& White 1997).} of the cluster is approximately $10^{15}$\,M$_{\odot}$.
%
%
%

We can also estimate the mass of the cluster from its Einstein radius:
\begin{equation}
\theta_{E}=\Big[ \frac{4GM(\theta_{E})}{c^{2}}\frac{D_{ls}}{D_{l}D_{s}} \Big]^{1/2}.
\label{ER}
\end{equation}
The Einstein radius is a robust measure of the lens mass
within this radius.
Although we did~not identify the counter-image(s)
of arc-A we can estimate the Einstein radius based on
the distance of the arc from the cluster bright galaxy ($\sim11''$).
Using Eq.~\ref{ER} we estimate that the total mass
enclosed within 90~kpc (i.e., the Einstein radius, $11''$,
at $z\approx1$) is $\sim6\times10^{13}$\,M$_{\odot}$.

To summarize, we report on the discovery of
SDSS\,120923.7$+$264047 -- a massive cluster of galaxies at $z=0.558$.
The cluster hosts many prominent arcs, including a giant arc with
$R$-band magnitude of 20.0, at $z=1.018$.
Moreover, presumably the cluster has a large Einstein radius,
suggesting that it is one of the richest strong
lensing clusters known.

\section*{ACKNOWLEDGMENTS}
Partial support for this work was provided by NASA through
grant HST-GO-10793.01-A from the Space Telescope Science Institute, which is
operated by the Association of Universities for Research in
Astronomy, Incorporated, under NASA contract NAS5-26555.
We thank Chuck Steidel for help with the spectrum analysis,
and Orly Gnat and Robert Quimby for comments on the manuscript.
Funding for the Sloan Digital Sky Survey (SDSS) and SDSS-II has been provided
by the Alfred P. Sloan Foundation, the Participating Institutions,
the National Science Foundation, the U.S. Department of Energy,
the National Aeronautics and Space Administration, the
Japanese Monbukagakusho, and the Max Planck Society,
and the Higher Education Funding Council for England.
The SDSS Web site is http://www.sdss.org/.

\end{document}